\documentclass[11pt]{article}
\usepackage{moriond,epsfig}
\usepackage{color}

\def\be{\begin{equation}}
\def\ee{\end{equation}}
\def\bea{\begin{eqnarray}}
\def\eea{\end{eqnarray}}

\begin{document}
\vspace*{4cm}
\title{An Improved Limit on the Electric Dipole Moment of the Muon}

\author{Ronald McNabb \\
(for the Muon g-2 collaboration)
}

\address{Dept. of Physics, University of Illinois at Urbana-Champaign \\
1110 W Green St., Urbana, IL 61801, USA.}

\maketitle\abstracts{
Data from the muon g-2 experiment at Brookhaven National Lab 
has been analyzed to search for a muon electric dipole 
moment(EDM), which would violate parity and time reversal 
symmetries.  An EDM would cause a tilt in the spin precession plane of 
the muons, resulting in a vertical oscillation in the position of 
electrons hitting the detectors.  No signal has been observed.  
Based on this analysis, an improved limit of 
$2.8 \times 10^{-19}$e-cm($95\%$ CL) is set on the muon EDM.
}

\section{Elementary particle EDMs}

  An EDM in an elementary particle would violate both parity and time 
reversal symmetries.  Since the spin is the only unique direction in 
an elementary particle, its EDM would have to be aligned with the spin 
direction.  The EDM changes sign under a parity inversion while spin remains 
unchanged.  Under time reversal the spin changes sign while the EDM 
is unaffected.  Assuming CPT invariance this implies that an elementary 
particle EDM would violate CP symmetry.

  Searches for an EDM have been performed with a number of elementary 
particles, but none has yet been observed.  A list of the current best 
limits for elementary particle EDMs is shown in table \ref{tab:EDMs}.
The two lowest limits are on the neutron and electron EDMs.  Both of these 
measurements set important constraints on CP violation beyond the standard 
model.
The current limit on the muon EDM is from the third CERN muon g-2 
experiment.  Since lepton EDMs scale linearly in most proposed 
theories and the electron EDM limit is many orders of magnitude lower, 
a muon EDM at this level would require some explanation for the lack 
of any observed electron EDM.  Thus the muon EDM limit has not been 
as important as the limit for the electron or neutron.

\begin{table}[ht]
\caption{The current best limits on elementary particle EDMs.}

\begin{tabular}{|c|c|c|} \hline

 Particle & Limit/Measurement (e-cm) & Method \\   \hline
e               & $<1.6 \times 10^{-27}$  &  Thallium beam \cite{Regan} \\ \hline
$\mu$           & $<1.05 \times 10^{-18}$ &  Tilt of precession plane in magnetic moment experiment \cite{CERN_EDM} \\ \hline
$\tau$          & $(-2.2 <d_{\tau}<4.5)\times 10^{-17}$ &  BELLE  $e^{+} e^{-} \rightarrow \tau \tau $ events \cite{Belle_tau} \\ \hline
n               & $<6.3 \times 10^{-26}$ & Ultra-cold neutrons \cite{Harris}  \\ \hline
p               & $(-3.7\pm 6.3) \times 10^{-23}$ & 120kHz thallium spin resonance  \cite{Proton_EDM}  \\ \hline
$\Lambda$       & $(-3.0\pm 7.4) \times 10^{-17}$ &  Tilt of precession plane in magnetic moment experiment \cite{Lambda_EDM} \\ \hline
$\nu_{e,\mu}$   & $<2 \times 10^{-21}$ &  Inferred from magnetic moment limits \cite{delAguila} \\ \hline
$\nu_{\tau}$    & $<5.2 \times 10^{-17}$ & Z decay width \cite{NuTau_EDM}  \\   \hline

\end{tabular}
\label{tab:EDMs}

\end{table}

\section{Experiment}
\subsection{Measurement of $a_{\mu}$ }

  The primary purpose of the muon g-2 experiment is to measure the muon 
anomalous magnetic moment($a_{\mu}$)\cite{g2_2001}.  In order 
to do this, 3.1GeV polarized muons are injected into a 7m radius storage 
ring with 
an extremely uniform magnetic field of 1.4T.  The muon spins precess in the
magnetic field with a frequency ($\vec \omega_{s}$) slightly higher than the 
cyclotron frequency ($\vec \omega_{c}$) of the muons.  The difference between 
these two frequencies is the g-2 frequency 
($\vec \omega_{a} = \vec \omega_{s}-\vec \omega_{c}$). 
This frequency is proportional to the anomalous magnetic moment of 
the muon:

\begin{equation}
\vec \omega_{a} = -\frac{e}{m_{\mu}c} a_{\mu} \vec B
\label{eq:precess}
\end{equation}

  The magnetic field is measured using fixed NMR probes placed around 
the ring and a trolley which moves through the storage region when the 
beam is off.  The anomalous precession frequency $\omega_{a}$
is measured by a set of 
24 calorimeters placed around the inside of the ring.  
Since electrons from muon decay are preferentially emitted along the 
direction of the muon spin, there is a difference in the rate of hits in
the calorimeters depending on whether the spins are oriented forward or 
backwards with respect to the muon spin direction.  So, as the muons 
precess, the rate of hits in the calorimeters
is modulated at frequency $\omega_{a}$.

\subsection{Effect of an EDM}

  If the muon has an non-zero EDM then the induced electric field in the 
muon rest frame would cause an additional torque on the muon.  This would
add a component to the spin precession orthogonal to both the vertical 
magnetic field and the direction of motion.  Equation \ref{eq:precess} 
then becomes

\begin{equation}
\vec\omega  = \vec \omega_{a} + {{\vec \omega_{EDM}}}
-\frac{e}{m_{\mu}c} \left[ a_{\mu}\vec B +
{{\frac{1}{2}f\left(\vec{\beta}\times\vec B\right)}}\right]
\label{eq:tilt}
\end{equation}

\noindent where $f$ is proportional to the muon EDM($D_{\mu}$)

\begin{equation}
D_{\mu}=f \frac{e\hbar}{4m_{\mu}c}
\label{eq:fdef}
\end{equation}

  As a result, the spin precession plane would tilt 
inward or outward with an angle $\delta \approx \frac{f}{2a_{\mu}}$.  
In addition the precession frequency would increase
($\frac{\Delta \omega}{\omega_{a}} = \frac{1}{2}\delta^{2}$) due to the 
EDM.  This could cause an error in the determination of the anomalous 
magnetic moment.  A muon EDM at the current limit of 
$<1.05 \times 10^{-18}$e-cm would cause a tilt of $9.3mrad$ and a 
shift in the precession frequency of 46ppm.

\begin{figure}[th] 

\hspace{1in}
\begin{picture}(200, 100)(-20,20)

\put(40.,20.){\scalebox{0.5}{\includegraphics{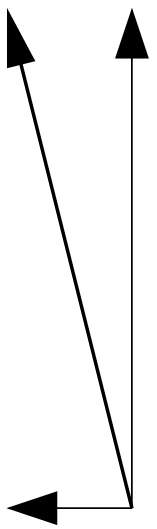}}}
\put(65, 90){$\vec{\omega}_{a}$}
\put(12, 20){${\vec{\omega}_{EDM}}$}
\put(20, 90){$\vec{\omega}$}
\put(52, 60){$\delta$}

\put(130.,20.){\scalebox{0.6}{\rotatebox{0}{\includegraphics{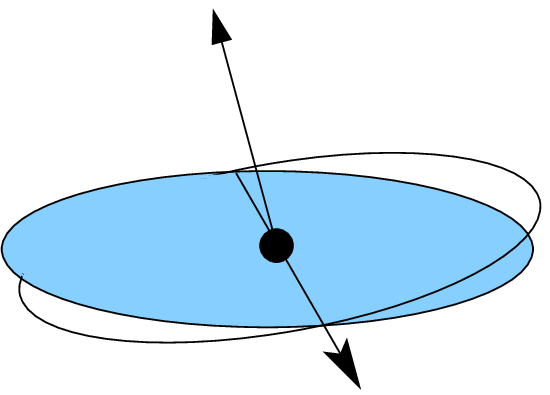}}}}
\put(158, 90){$\vec{\omega}$}
\put(195, 15){$\vec{\beta}$}

\put(230.,15.){\scalebox{0.2}{\includegraphics{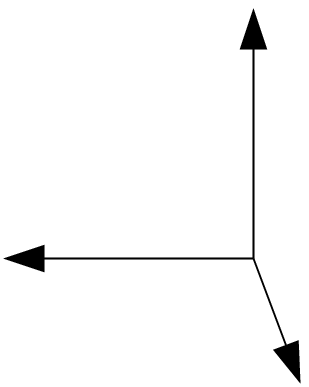}}}
\put(250, 32){\tiny $\vec{B} $}
\put(250, 15){\tiny $\vec{\beta}$}
\end{picture}

\caption{A muon EDM would tilt the spin precession plane.
}

\label{fig:spintilt}
\end{figure}

\subsection{Signal for an EDM}

  A tilt in the precession plane would cause an oscillation of the 
vertical muon spin component at $\omega_{a}$, but $90^{\circ}$ out of 
phase with the oscillation in rate which is used to measure the 
precession frequency.  Since electrons from muon decay are 
preferentially emitted in the direction of the spin the combined precession 
would cause 
an oscillation in the average vertical angle of emitted electrons.  As
a result, electrons hitting the detectors would show a vertical oscillation
in their detected positions.  Detailed GEANT simulation gives the expected 
size of oscillations for a given EDM as $(8.8\pm0.5)\mu m$ per 
$10^{-19}$e-cm.  An EDM at the current limit would cause a 
$90\mu m$ oscillation.

  The positions of electrons hitting the calorimeters were measured by 
a set of five scintillating tiles which cover the front face of the
calorimeters, providing vertical segmentation.  This is referred to as 
the Front Scintillator Detector(FSD).  In the 2000 g-2 data run, nine
calorimeters were outfitted with FSDs.  Data from these 
detectors has been used to search for a vertical oscillation in the 
position of hits on the calorimeters.

\begin{figure}[th] 
\begin{center}
\includegraphics[width=.29\textwidth]{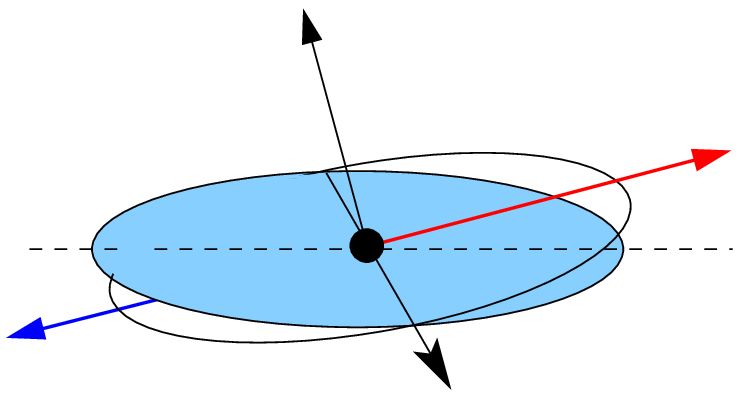}
\hspace{0.3cm}\rule{0.01mm}{35mm}\hspace{0.3cm}
\includegraphics[width=.59\textwidth]{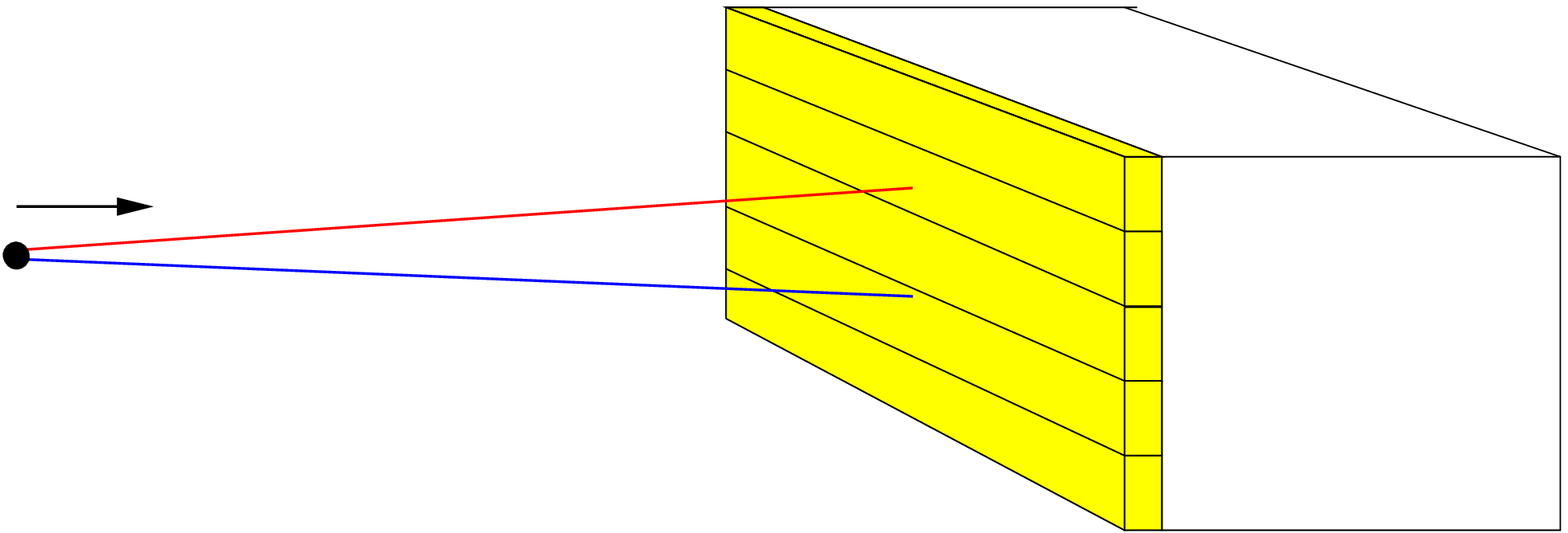}
\end{center}
\caption{A tilt in the precession plane results in a vertical 
oscillation of hits on the detector face.
}

\label{fig:FSD}
\end{figure}

\section{Analysis}
\subsection{Position versus Time}

\begin{figure}[th] 
\begin{center}
\includegraphics[width=.89\textwidth]{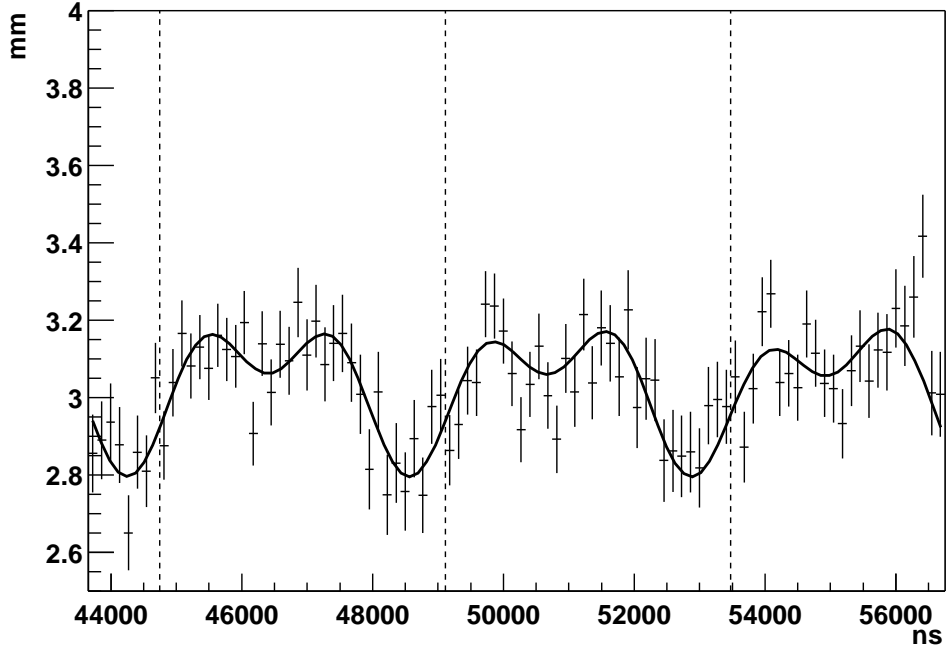}
\end{center}
\caption{Mean position of hits on a detector versus time.  The
dashed lines indicate the zero crossings for the rate versus time
plot.}
\label{fig:yvst}
\end{figure}

  First, hits in the FSDs are matched with corresponding hits in the 
calorimeters.  Thus, the time, position, and energy of each event can 
be determined.  Electrons with energies between 1.4GeV and 3.2GeV
are accepted. The lower cut is just above the triggering threshold 
for the calorimeters and the upper cut is at the endpoint of the 
electron energy spectrum.  Then, the mean position of hits on the FSD
is plotted versus the time since the muons were injected.  A example 
of data from a single detector, in a short time span, is shown in figure 
\ref{fig:yvst}.

  The plot shows two oscillations, with the longer period oscillation at
the g-2 frequency.  The second component is the coherent betatron 
oscillation(CBO) frequency, which describes the collective radial motion 
of the muon beam.  The data is fit to:

\begin{eqnarray}
 y(t) &=& \textcolor{red}{Y_{0}} + \left[\textcolor{red}{S_{g2}}\,sin\left(\omega_{a}t\right) + \textcolor{red}{C_{g2}}\,cos\left(\omega_{a}t \right)\right]   \nonumber \\ \nonumber \\
 &+& e^{-\frac{t}{\tau_{c}}} \times  \left[
 \textcolor{red}{{S}_{c}}sin\left(\omega_{c}t\right) +  \textcolor{red}{{C}_{c}}cos\left(\omega_{c}t \right) \right]  \nonumber \nonumber
 \end{eqnarray}

\noindent where the free parameters are the sine and cosine coefficients
and an overall offset.  The g-2 oscillation is phase aligned so that any
EDM signal would appear in the sine term of the g-2 frequency oscillation,
$90^{\circ}$ out of phase with the rate oscillation.  The CBO frequency 
term decays with a $150\mu s$ time constant because of dephasing due to 
the frequency spread of the CBO.

\subsection{The EDM Signal}

\begin{figure}[p] 
\begin{center}
\includegraphics[width=.89\textwidth]{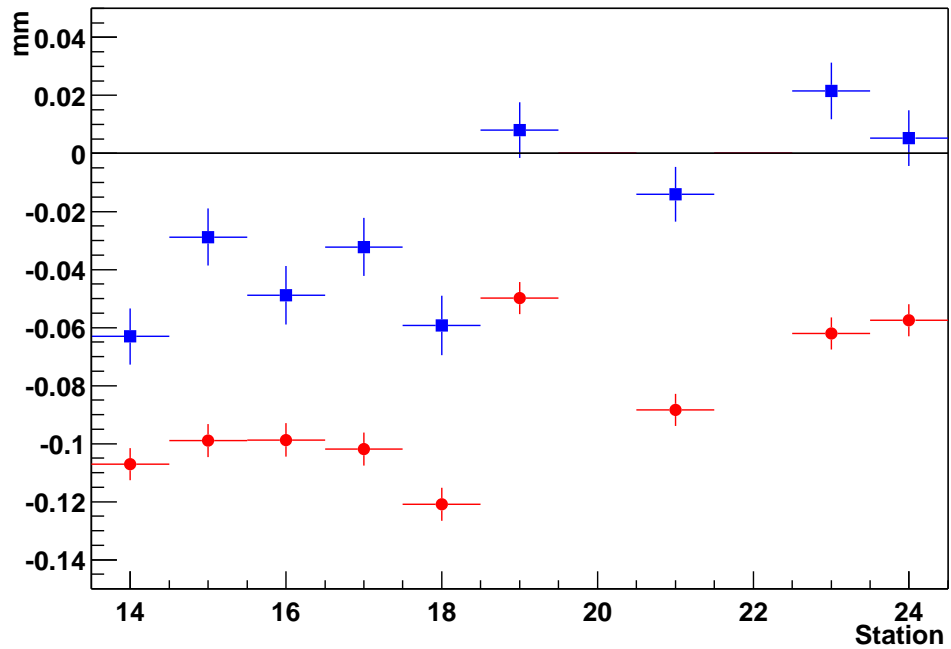}
\end{center}
\caption{Coefficient of the g-2 sine term (the EDM signal) versus station 
number.  The red circles are data from before the beam realignment, the 
blue squares are data from after.
}
\label{fig:svsstat}
\end{figure}

  Figure \ref{fig:svsstat} shows the fitted coefficient of the g-2 sine 
term versus detector number.  As noted, this is the term where any EDM
induced oscillation would occur.  The data is split into two sets, since 
about $\frac{2}{3}$ of the way through data taking the muon beam was moved
vertically to better align it with the detectors.  Obviously, there is a 
difference in the EDM signal between data before(red circles) and 
after(blue squares) the 
realignment.  Also, there are unacceptable variations in the signal from 
detector to detector.

  These variations are not unexpected since there is a known systematic 
effect due to 
the alignment of the beam and detectors.  Electrons emitted radially 
outward take longer to travel to the detectors than those emitted inward.
A longer time of flight allows the electron vertical profile to spread 
more before hitting the detector.  Thus, as the muon spins precess,
there is an oscillation in the width of the vertical profile at the g-2 
frequency, in phase with the EDM signal.  This is not a problem if the 
detector is perfectly aligned with the beam.  However, if the beam is offset
upward, then as the width changes more electrons are lost off the top 
edge of the detector when the beam spreads out.  Asymmetric losses result 
in an oscillation of the measured mean vertical position.  

  The vertical alignment was the primary 
systematic error in the previous muon g-2 experiment's EDM measurement.
Unfortunately, the beam to detector alignment is only known to within
approximately 0.5mm, similar to the knowledge from the previous 
experiment.  So, in order to improve the measurement, a new analysis 
technique is necessary.

\subsection{Eliminating the Offset Error with CBO}

\begin{figure}[p] 
\begin{center}
\includegraphics[width=.89\textwidth]{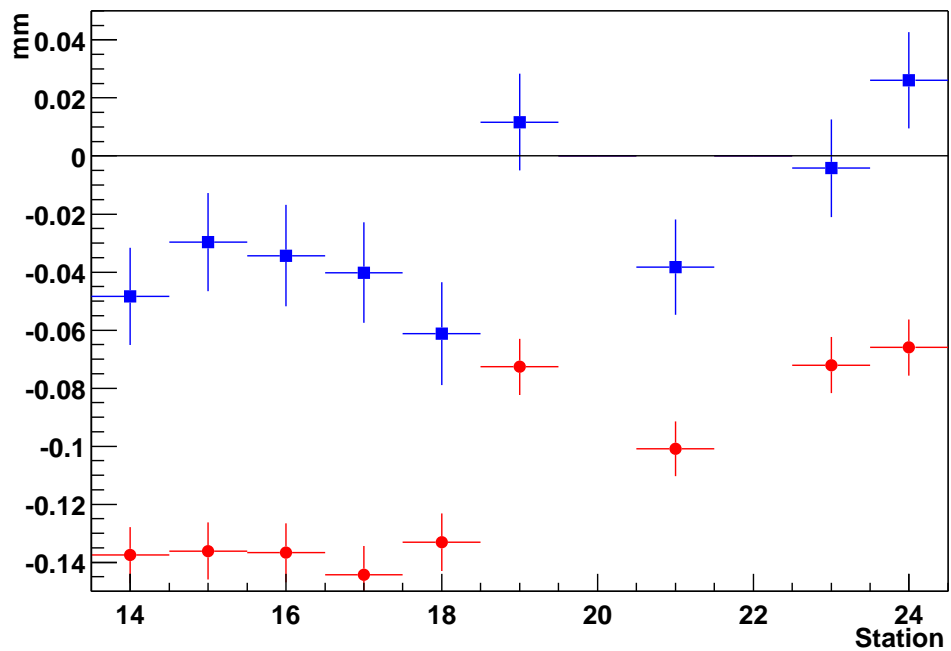}
\end{center}
\caption{Coefficient of the CBO sine term versus station 
number.  The red circles are data from before the beam realignment, the 
blue squares are data from after.
}
\label{fig:svsstatc}
\end{figure}

  As the beam oscillates radially with the CBO frequency the time of 
flight to the detectors changes.  It is longer when the beam is further 
out radially.  Thus, there is an oscillation in the width of the profile
at the CBO frequency, similar to that seen at the g-2 frequency.  This 
would not result in any oscillation in the mean versus time if 
the detector 
were perfectly centered on the beam.  However, with a misalignment, there 
are more losses off of one edge of the detector as the beam spreads, 
resulting in an oscillation of the measured mean.  This is the same effect
that occurs at the g-2 frequency.  The plot in figure \ref{fig:svsstatc}
shows the CBO frequency oscillation amplitude versus detector from the 
same fits as the g-2 oscillation amplitude.  Again, data from before the 
beam alignment shows a much larger oscillation amplitude.

\begin{figure}[p] 
\begin{center}
\includegraphics[width=.89\textwidth]{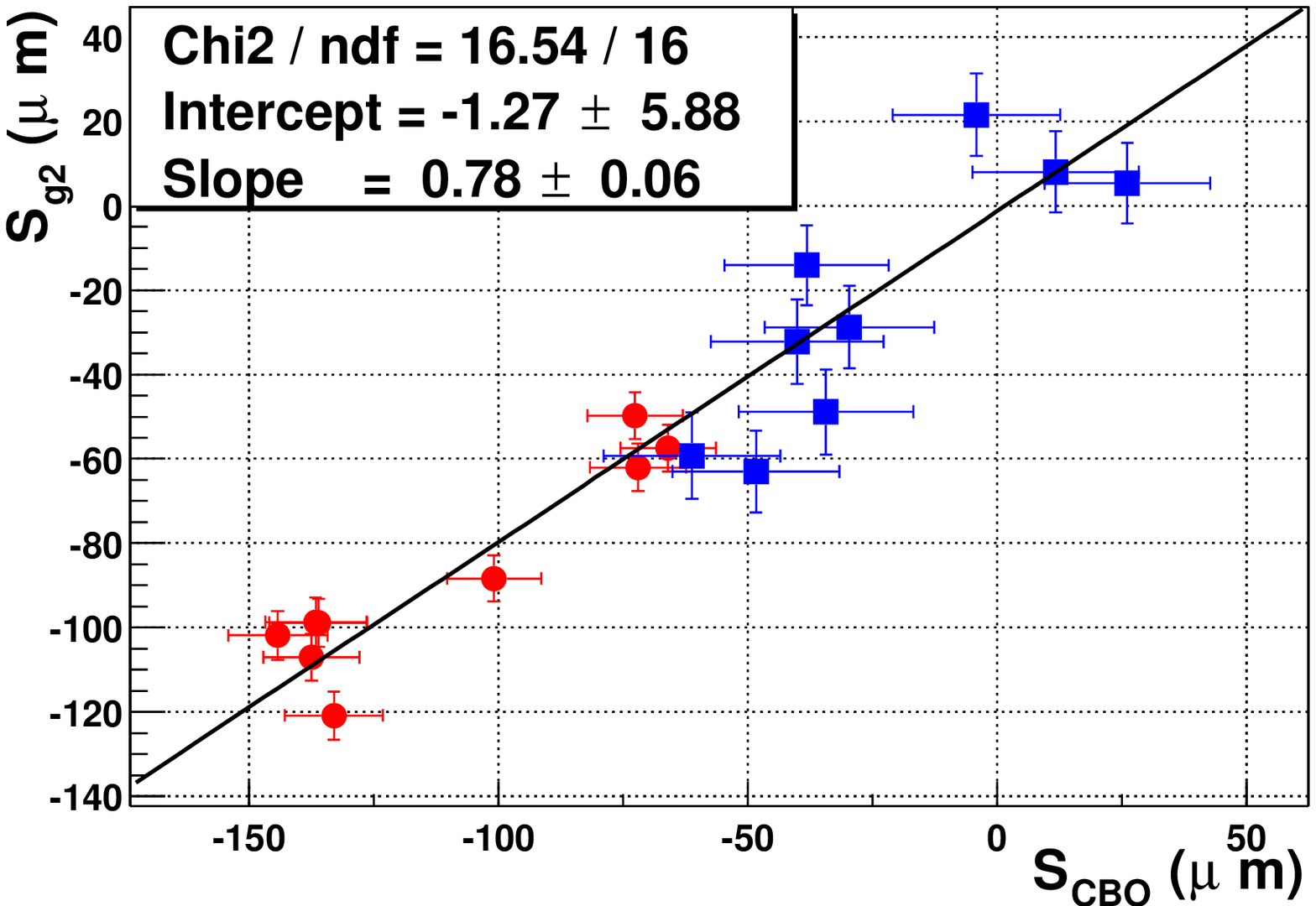}
\end{center}
\caption{The g-2 sine amplitude versus the CBO frequency amplitude for each
of the 9 detectors before(red circles) and after(blue squares) the beam 
realignment.
}
\label{fig:g2vscbo}
\end{figure}

  Since the oscillation in the mean at the CBO frequency is caused by 
the vertical offset of the detector, a detector that has no oscillation 
is aligned and should show no systematic effect due to the alignment at 
the g-2 frequency.  Figure \ref{fig:g2vscbo} shows a plot of the 
g-2 sine phase amplitude and the CBO amplitude from the plots in figures
\ref{fig:svsstat} and \ref{fig:svsstatc}, for
each of the nine detectors.  Again, data from before the realignment is 
represented by red circles and data from after is represented by blue
squares.  The points are fit to a straight 
line which gives a good $\chi^{2}$. 

  The point where there is no oscillation at the CBO frequency corresponds
to the point at which there is no effect due to offset at either frequency.
Thus, the intercept of the fit line corresponds to the EDM measurement, 
corrected for the effect of detector offset.  The intercept is 
$-1.27 \pm 5.88\mu m$.  Comparing this to the simulation mentioned above
gives an EDM of  $(-0.14 \pm 0.67)\times 10^{-19}$e-cm, with only 
the statistical error.

\section{Systematic Errors}

  Table \ref{tab:syst} shows the systematic errors in the analysis.
The detector tilt error results from the horizontal oscillation of the
profile on the calorimeter face.  A horizontal oscillation combined with 
a tilt in the detector would cause an apparent 
vertical oscillation.  The average detector tilt was measured to be less 
than $\frac{1}{2}^{\circ}$.  This tilt is combined with the amplitude of 
horizontal oscillations, from simulation, to obtain a systematic uncertainty.

  The vertical spin systematic error stems from the fact that, if the muons
have a non-zero vertical polarization, the electrons will have a  
vertical angle at which they are emitted.  Since there are oscillations 
in the path length with time, this would cause a vertical oscillation on 
the detector face.  This systematic error is determined through data
from the traceback chamber, which consists of a set of straw tube 
hodoscopes that measure the angle of the electrons as they approach the 
detector.

  A tilt in the electrostatic quadrupoles, which focus the beam in the 
ring would cause a vertical oscillation in the beam position at the CBO 
frequency.  The limit on the quadrupole tilt is obtained from surveys 
of the quadrupole plates.  The tilt is combined with an estimate of the CBO 
horizontal oscillation amplitude to obtain a systematic error.

  The top and bottom of the calorimeter are read out by different 
photomultiplier tubes.  These tubes my have different gains and 
timing offsets and there is the potential for a systematic error 
due to this vertical asymmetry.  The systematic error estimates
for these effects are obtained by artificially inflating the effects 
in software and rerunning the analysis to look for changes in the 
result.  
     
\begin{table}[p]
\begin{center}
\caption{Systematic Error Table}

\begin{tabular}{|c|c|} \hline

Error   &  $\mu m$ \\
\hline\hline

Detector Tilt &  6.1    \\
\hline

Vertical Spin   &  5.1     \\
\hline

Quadrupole  Tilt   & 3.9          \\
\hline

Timing Offset & 3.2       \\
\hline

Energy Calibration   & 2.8   \\
\hline

Radial Magnetic Field   & 2.5   \\
\hline

Albedo and Doubles & 2.0   \\
\hline

Fitting Method & 1.0   \\
\hline

\hline

Total Systematic        & 10.4      \\

\hline
Statistical & 5.9   \\
\hline
Total Error &  11.9  \\
\hline
\end{tabular}

\label{tab:syst}
\end{center}

\end{table}

\section{Results}

  The systematic errors above are all uncorrelated, thus they are 
added in quadrature to obtain a total systematic error of $10.4\mu m$ 
on the oscillation amplitude.  The systematic error is combined in 
quadrature with the statistical 
error of $5.9\mu m$ to obtain a total error of  $11.9\mu m$.  Based on 
this, the EDM measurement with statistical and systematic errors 
combined is $(-0.1 \pm 1.4)\times 10^{-19}$e-cm.  
Since the result is consistent with zero a $95\%$ confidence level 
limit of $|D_{\mu}| \leq 2.8 \times 10^{-19}$e-cm is set,
a factor of 4 improvement over the current limit.

\end{document}